\documentclass[12 pt]{article}        	
\usepackage{amsfonts, amssymb}
\usepackage{amsmath}
\usepackage{graphicx}
\usepackage{float}
\usepackage{esint}
\usepackage[left=2cm,right=2cm,top=2cm,bottom=2cm]{geometry}
\newcommand{\qed}[0]{$\square$}

\newcommand{\z}{\mathbb Z}
\newcommand{\real}{\mathbb R}

\newcommand{\n}{\mathbb N}

\usepackage{units}
\newtheorem{theorem}{Theorem}
\newtheorem{lemma}{Lemma}

\title{A Generalized Benford Framework for Threat Identification in Counter-Intelligence}

\date{January 2025}

\begin{document}

\maketitle
\begin{center}
    \textbf{Author}

    Timothy Tarter, James Madison University

    Department of Mathematics

    www.tartermathematics.com
    
\end{center}

$\newline$

\begin{abstract}
    In this paper, we develop a framework of `Benford models' for counter-intelligence investigations which analyze frequency data of a suspect's visits to physical locations, online websites, and communication channels. We accomplish this by establishing the Benford measure for continuous \& bounded domains, generalizing the accumulated percentage differences between sites in the frequency data with the log-determinant of `Benford Matrices,' employing an estimator to determine a `Benford Test Statistic ($\lambda$),' and identifying maximal values of $\lambda$ across all permutations of included sites in our data. This framework is intended to complement outlier analysis models by finding where hidden Benford patterns `break' in frequency data and telling investigators which sites they should investigate.
\end{abstract}

\section{Overview}

Before reading this paper, please take the following space below (or a nearby piece of paper) and draw seven X's at random. 

\begin{figure}[H]
    \centering
    \includegraphics[width=0.75\linewidth]{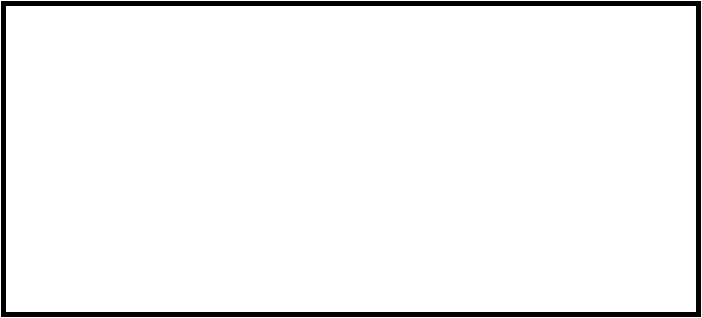}
    \label{fig:enter-label}
\end{figure}

$\newline$
Without seeing your page in person, the author can tell you that the distance between a given point and each closest point is about the same between all points. To help illustrate this, see figure 1:

\begin{figure}[H]
    \centering
    \includegraphics[width=0.5\linewidth]{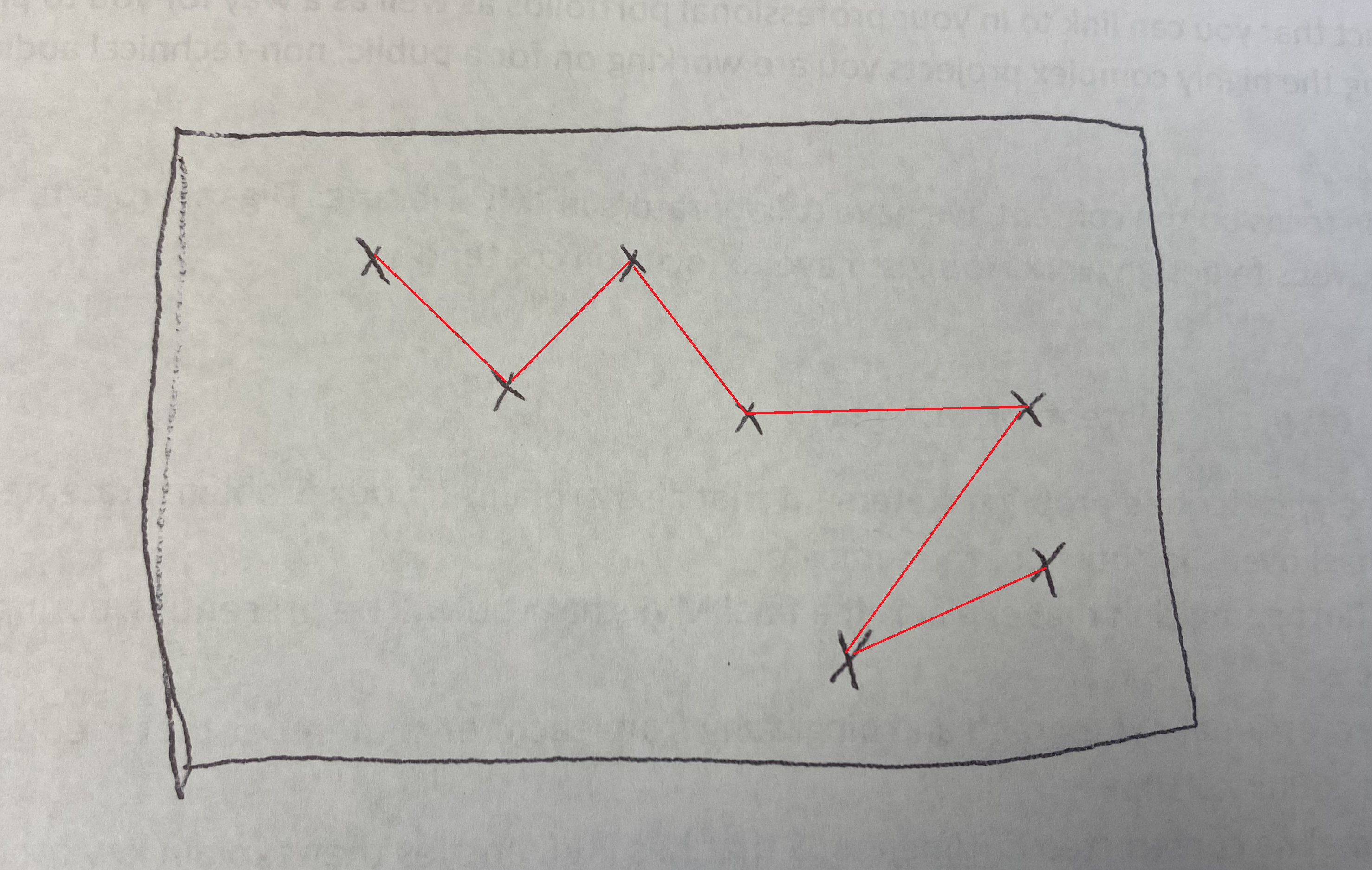}
    \caption{Demonstration}
    \label{fig:enter-label}
\end{figure}

$\newline$
The key word in the prompt provided above was ``random." The human brain is notoriously poor at making random patterns when it is trying to (Małgorzata et. Al., 2008). This is one reflection of Benford's Law which describes a surprising relationship between data randomly produced by natural processes and the probability distribution of leading digits in base-invariant representation. In the past, Benford's Law has been used to detect financial fraud because the property is surprising to most individuals attempting to fabricate data. 

\paragraph{Law of Anomalous Numbers}
A very familiar probability model examines an intuitive notion: if we are given nine items and one slot to put one of them in, the probability that a given number ends up in the slot is $\frac{1}{9}$ - i.e. equilikely. If we generalize this principle to nine numbers, $\{1, 2, 3, 4, 5, 6, 7, 8, 9\}$, we would assume that for any given slot, the distribution of numbers would be equilikely. To his surprise, however, in 1881, Simon Newcomb (and later Frank Benford) noticed that the leading digit in table entries of various datasets was more often one than two, two than three, three than four, \textit{et cetera}. 

\begin{figure}[H]
    \centering
    \includegraphics[width=0.4\linewidth]{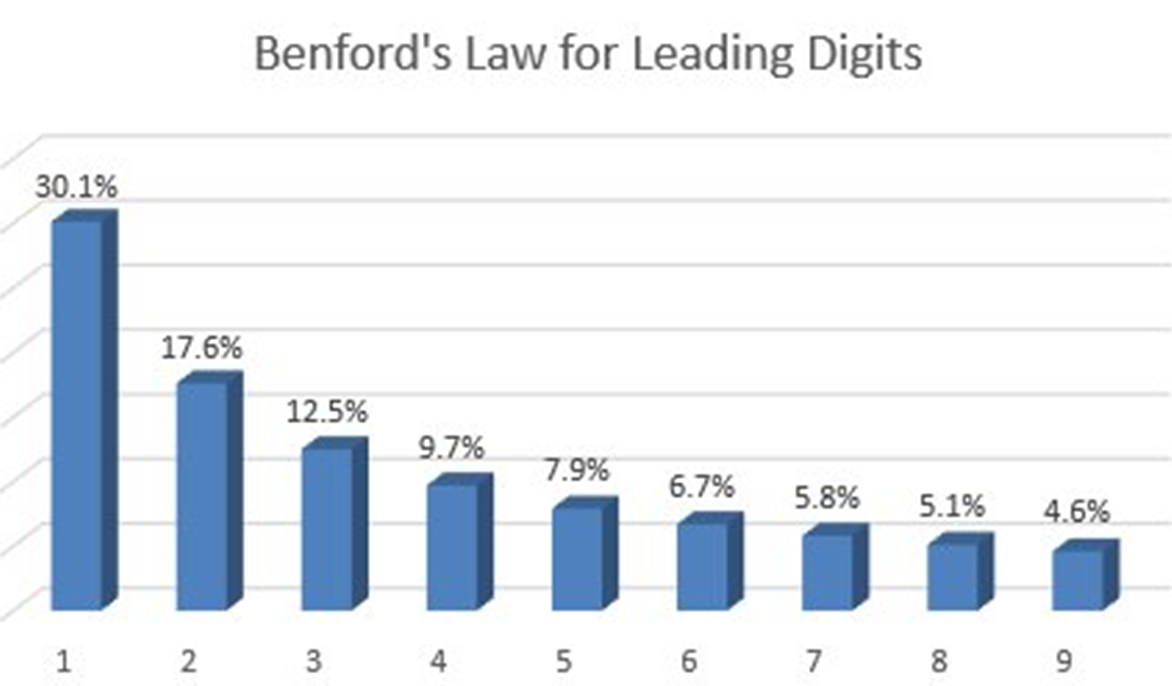}
    \caption{Law of Anomalous Numbers Visualization (Frost, 2022)}
    \label{fig:enter-label}
\end{figure}

\paragraph{Early Applications}
Slightly after the discovery of base-invariance and implications of goodness of fit of Benford's distribution across orders of magnitude (Miller, 2017), Benford's Law was used to unearth Bernie Madoff's ponzie scheme. Since then, ``Benford Tests" have been used in financial crime investigations as a way to catch criminals altering bank statements, tax forms, and other types of money-laundering schemes. 

\section{Defining the Benford Measure}
\paragraph{The Discrete Characterization of Benford Measure}
The Benford Distribution, denoted $\bold{B}$, is the unique measure on ($\real^+$, S), where S is some leading digit (or for our purposes, generalized distance), such that the accumulation of leading digits [1, t] follows a logarithmic distribution (for some base $k$, which is typically 10): 
\begin{equation}
    \bold{B} = \mu(\bigcup_{\ell \in \z} k^{\ell} [1,t]) \simeq ln(t), t \in \n_{[k]}
\end{equation}
Moreover, its base-10 PDF is defined by:
\begin{equation}
    P(S \leq t) =  log_{10}(t+1)-log_{10}(t), \forall t \in [1,9]
\end{equation}
\paragraph{Generalization of Benford Measure to a Continuous Domain}
This generalization is useful for continuous domains under a distance metric. It is also unique on its domain and carries the same properties as the above measure on $\real^+$ (Miller, 2017). 

$\newline$
We previously established the discrete PDF for leading digits with a log base 10. Using the base invariance property of the distribution established by previous literature, we find that a generalized discrete Benford measure for base $k$ is 
\begin{equation}
    P(S \leq t) = log_{k}(1+\frac{1}{t}), \forall t \in [1,k-1].
\end{equation}
We now want to motivate two things:
\begin{enumerate}
    \item A continuous domain for t
    \item Moving $log_k$ to $log_e$
\end{enumerate}
Motivating the log-base $e$ is fairly simple. Since the original measure follows from the accumulation of $\frac{1}{t}$, we can let $a$ be in the neighborhood of some t $\in$ [1,k]. Then the image of $a$ through the map:
\begin{equation}
    f:X \to Y, X \in (0,1], Y \in (-\infty, ln(k)]
\end{equation}

$\newline$
will be in the neighborhood of the image of $t$. Additionally, let $d$ denote a distance metric on both X and Y. Formally, letting $\epsilon$ $>$ 0, if a $\in$ $n_{\epsilon}(t)$ then $\exists$ $\delta$ $>$ 0 such that f(a) $\in$ $n_{\delta}(f(t))$. Moreover, since $\delta$ $\in$ $\real^+$, by the Archimedean property, we find that $\forall$ a,t $\in$ X, $\exists$ K $\in$ $\real^{\geq 0}$ with d(f($x_1$), f($x_2$)) $\leq$ K d($x_1$,$x_2$). This means that $f$ is Lipschitz continuous. Therefore, as the domain of X expands (moving the base of our distribution towards $\infty$),  our property of Lipschitz continuity allows us to express an approximation of that continuous distribution with log-base $e$.

$\newline$Now, we want to justify t $\in$ $\real^+$, not just $\z^+$. For this argument, we consider $\bold{B}$ as an approximation of $ln(t)$. While the measure was defined this way for the aforementioned purpose of analyzing discrete values, we find that moving t to a continuous domain in $\real$ is analogous to the Riemann sum converging to a definite integral. Thus, we introduce the measure
\begin{equation}
    \bold{\Tilde{B}} = \bigcup_{k\in\z} 10^k [1,t] = ln(t), t \in \real^+
\end{equation}
With a continuity approximation error term of O($\bold{B} - \bold{\Tilde{B}}$). Our PDF becomes,
\begin{equation}
    P(S \leq t) = ln(1 + \frac{1}{t}), \forall t \in [\frac{1}{e-1}, \infty).
\end{equation}
Again, recall that the justification for the bounds of the distribution comes from our argument of Lipschitz continuity and nearness of fit to continuum-normalized distributions. Making a simple substitution of variables, x = 1 + $\frac{1}{t}$, we obtain a ``better behaved" distribution:
\begin{equation}
    P(S \leq x) = ln(x), x \in [1,e]
\end{equation}

\section{Counter-Intelligence Threat Identification}

This research started because of a thought experiment about the famous spy, Robert Hanssen. In 1976, Hanssen was hired as an FBI agent and quickly promoted to a position in the FBI's counter-intelligence department. Between 1979 and 2001, Hanssen played turncoat, providing Soviet and Russian intelligence agencies with secrets vital to the state of American National Security. Arguably the most alarming aspect of Hanssen's tale is that he managed to do this while under near-constant surveillance. This highlights certain key flaws of otherwise powerful forms of outlier analysis for counter-intelligence investigations: one must assume that outliers will occur before substantial damage is done. Hanssen's story prompted the question: given similar circumstances (constant surveillance via geo-spatial tagging, monitoring communication, etc.), how can we prevent Hanssen-esque threats in the future? Specifically, can Benford models offer promising solutions by highlighting subtle, hidden patterns in data?

\subsection{Frequency Analysis}
To specify a use case for our solution, we want to narrow down the types of data that we intend to use to frequency data about the number of times our suspect visits a set of locations over a certain period of time. This is powerful because it easily generalizes from a suspect visiting physical locations to the communication habits, online presence, and spending patterns of a suspect. More specifically, however, we want to compare the percent change between the frequency counts at each site, the same way that we compared distance in [1]. 

\begin{lemma}
    Let $f_i$ denote the number of visits to a site i, letting n be the total possible number of sites. Then, 
    $g_{i,j} \simeq ln(f_i) - ln(f_{j})$ is the percent change between $f_i$ and $f_{j}$.
\end{lemma}
Proof (Sims - Math Review, 2015):
\begin{itemize}
    \item We know the relative difference frequency, $g_i$ = $\frac{f_i - f_{j}}{f_{j}}$ = $\frac{f_i}{f_{j}} - 1$
    \item Then $\frac{f_i}{f_{j}} = g_{i,j} + 1$
    \item ln($g_i + 1$) = ln($f_i$) - ln($f_{j}$) $\simeq$ $g_{i,j}$
\end{itemize}
\hfill \qed

\subsection{Order Invariance of Sites \& Generalizing $\mu$}
We use this to compare percentage changes between visits at sites using $ln(\frac{f_i}{f_j})$. That said, in the real world, we aren't just looking at two sites, and there isn't an intrinsically correct ordering (as opposed to time-series data for example). This motivates us to want to generalize the percentage change in each site with respect to every other site via \textbf{Benford Matrices}.

$\newline$
Consider a trivial example where there are three sites which yield the 3 x 3 \textbf{Benford Matrix}:

$\newline$
\begin{equation}
    A = \begin{bmatrix}
    \nicefrac{f_1}{f_1} & \nicefrac{f_2}{f_2} & \nicefrac{f_3}{f_3} \\
    \nicefrac{f_1}{f_2} & \nicefrac{f_2}{f_3} & \nicefrac{f_3}{f_1} \\
    \nicefrac{f_1}{f_3} & \nicefrac{f_2}{f_1} & \nicefrac{f_3}{f_2}
\end{bmatrix}
\end{equation}

$\newline$

$\newline$Aside: for `n' sites, this will be an n x n matrix.

$\newline$
Then to capture the total percentage of differences between sites, $\Delta f$, let ln$|det(A)|$ = $\Delta f$ (since the determinant generalizes the accumulated area of a collection of vectors). Additionally, this is justified by the fact that multiples of Benford variables and their reciprocals are also Benford (Pike, 2008). Since the absolute value of a determinant divided by the number of vectors in its matrix tells you the average scaled volume of its columnspace (i.e. how spread out the vectors are), our formula will tell us how spread out the patterns in our data are. If that spread is similar to a ``Benford spread," then the patterns in our suspect's data are Benford patterns. If they aren't similar, then there is an outlier to the expected Benford pattern for natural processes. Note that the expected average scaled volume is the expected value of a Benford variable, which we have previously established to be $\sim$2.0973. This is a particularly nice generalization for the following reason: we can reasonably expect the total percentage of differences between sites to be equal to the expected value of the continuous Benford distribution if and only if the true frequency data is Benford. 
Thus, for a continuous random variable, B $\sim$ Benford, and for n sites of $f$,
\begin{theorem}
    $E[\mathbf{B}] = \frac{\Delta f}{n}$ = $\frac{ln|det(A)|}{n}$ iff $\bigcup_{i}f_i$ $\sim$ Benford
\end{theorem}

$\newline$
Note: using our ``well-behaved" generalization of Benford Measure to a continuous (but bounded) domain in (7), with X a continuous random Benford variable, $f_X(x) = ln(x)$, we derive:
\begin{equation}
    E[X] = \int_{1}^{e} x ln(x) dx \simeq 2.0973 + O(\bold{B} - \bold{\Tilde{B}})
\end{equation}

\subsection{The Benford Test Statistic}
Given the nature of the intended application for this framework, we know that this parity relation will very rarely be held. In fact, because we want to know where to advise investigators to ``look" for illicit activity from a suspect, it may even be more useful to study the magnitude of perturbation caused by introducing a site to the dataset than merely noting that the parity relation is broken. Therefore, let us employ an estimator, $\lambda$, our \textbf{Benford Test Statistic}, to tell us how broken our parity relation is. 

\begin{equation}
    \lambda = 2.0973 - \frac{ln|det(A)|}{n}
\end{equation}

$\newline$
It is worth noting that if the sample data is Benford, then $\lambda$ will be in some close neighborhood of zero. Since we assume our sample to be Benford, then the radius of that neighborhood in $\real^1$ will be the standard deviation of the Benford distribution for one variable.

\subsubsection{Higher Moments of the Benford Distribution}
We calculate these higher moments using our distribution $\bold{\Tilde{B}}$ from (7):
\begin{equation}
    E[X] = \int_{1}^{e} x ln(x) dx \simeq 2.0973 + O(\bold{B} - \bold{\Tilde{B}})
\end{equation}

\begin{equation}
    E[X^2] = \int_{1}^{e} x^2 ln(x) dx \simeq 4.5746 + O(\bold{B} - \bold{\Tilde{B}})
\end{equation}

\begin{equation}
    V[X] = 4.5746 - 2.0973^2 \simeq 0.1759 
\end{equation}

\begin{equation}
    \sigma[X] = \sqrt{0.1759} \simeq 0.4149
\end{equation}

\subsubsection{Hypothesis Testing Formula}
From these higher moments, we can perform hypothesis tests in the following manner:

$\newline$$H_0$: $\lambda$ = 0; $H_A$: $\lambda$ $\not=$ 0

$\newline$
Test Statistic (Studentized t-test):
\begin{equation}
    t = \frac{0-\lambda}{0.4149}
\end{equation}

$\newline$Our p-value can be calculated with df (degrees of freedom) = $n^2$ - 1 (since our Benford Matrix is n x n), and can be reasonably evaluated at the 95\% level of significance.

\subsection{Maximizing $\lambda$}
Recalling our motivation for finding $\lambda$ in [3.3], we really want to determine which entries $a_{i,j}$ $\in$ A maximize $\lambda$ when added to our Benford Matrix, A. Since our sites are order-invariant, we unfortunately have to select out one column at a time (which of course also removes a row, making A (n - 1) $\cdot$ (n - 1)), which is somewhat computationally expensive. Moreover, it isn't enough to just do this for one removed row / column, we have to do it for every combination of sites which could be removed from the dataset. Thus, for `n' sites, we will have to run n$\cdot$n! computations of $\lambda$, and then employ an optimization algorithm to find $\max_{i\cdot j}\lambda_{i,j}$. Simplifying this permutation algorithm is classified under active future research.

\subsection{Numerical Simulation in Python}
To validate our Benford Framework for frequency analysis, we programmed numerical simulations in Python which sample from various distributions and perform the steps detailed above for Benford our tests.

$\newline$
If the reader is so inclined, they can access the code on my website at:

\begin{center}
    www.tartermathematics.com
\end{center}
Otherwise, here are some screenshots of the example outputs from the code:
\begin{figure}[H]
    \centering
    \includegraphics[width=0.75\linewidth]{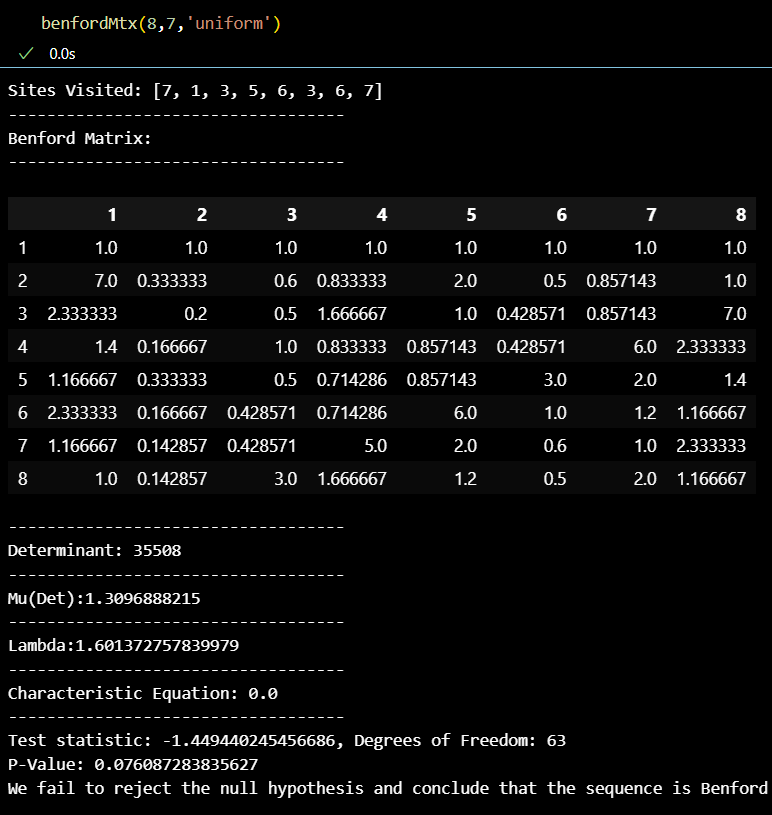}
    \caption{Sample Benford Set}
    \label{fig:enter-label}
\end{figure}
\begin{figure}[H]
    \centering
    \includegraphics[width=0.75\linewidth]{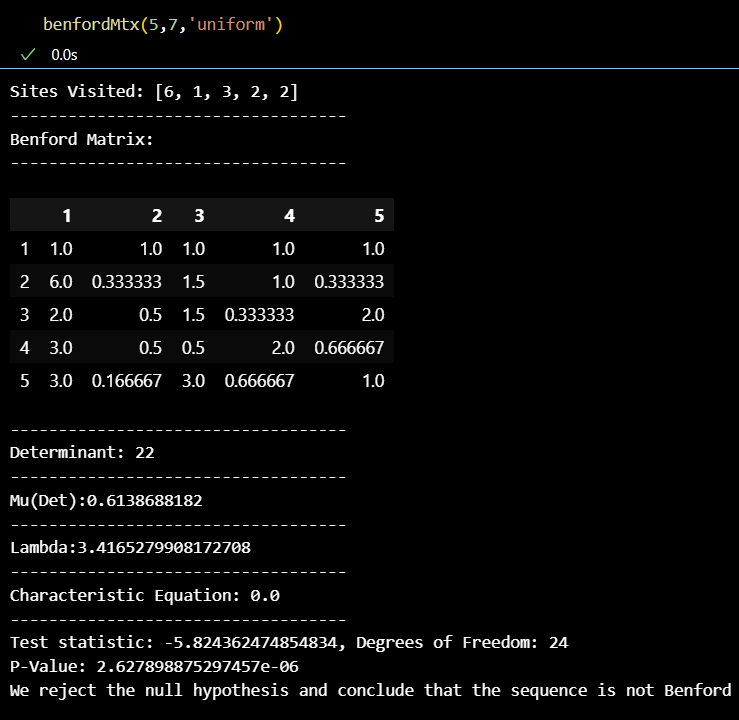}
    \caption{Sample `Not' Benford Set}
    \label{fig:enter-label}
\end{figure}

\section{Conclusion}
In this work, we generalize the Benford Measure beyond the Significand $\sigma$-Algebra, characterize a continuous distribution, as well as a bijection from it to a bounded continuous distribution, establish an algorithm for determining if a sequence of frequency data is distributed Benford, and provide an algorithm for analyzing system perturbations. These results indicate likely highly material performance yields for future research on Benford models, especially for applications in national security and threat prevention.

\vfill

\section{References}
Frost, J. (2022, October 6). Benford’s Law explained with examples. Statistics By Jim. 

https://statisticsbyjim.com/probability/benfords-law/ 

$\newline$
Park, S.-H., Huh, S.-Y., Oh, W., \& Han, S. P. (2012). A Social Network-Based Inference Model for Validating Customer Profile Data. MIS Quarterly, 36(4), 1217–1237. 

https://doi.org/10.2307/41703505

$\newline$
Morzy, Mikolaj \& Kajdanowicz, Tomasz \& Szymanski, Boleslaw. (2016). Benford’s Distribution in Complex Networks. Scientific Reports. 6. 34917. 10.1038/srep34917. 

$\newline$
Miller, S. J. (2017). Benford’s law: Theory and applications. Princeton University Press. 

$\newline$
(PDF) Benford’s law and the C$\beta$e. (n.d.). 

https://www.researchgate.net/publication/368304741\_Benford’s\_law\_and\_the\_CbetaE

$\newline$
Kopczewska, K., \& Kopczewski, T. (2022, October 20). Natural spatial pattern-when mutual socio-Geo Distances between cities follow Benford’s law. PloS one. 

https://pmc.ncbi.nlm.nih.gov/articles/PMC9584388/ 

$\newline$
Author links open overlay panel Małgorzata Figurska a, a, b, 1, believed, S. is widely, Brown, R. G., Daniels, C., Joppich, G., Persaud, N., Schneider, S., Pollux, P. M. J., \& Baddeley, A. D. (2007, September 20). Humans cannot consciously generate random numbers sequences: Polemic study. Medical Hypotheses. 

https://www.sciencedirect.com/science/article/abs/pii/S030698770700480X 

$\newline$
Pike, D. (2008). Testing for the benford property. SIAM. 

https://www.siam.org/media/r3alxzk0/testing\_for\_the\_benford\_property.pdf 

$\newline$
Friar, J. L., Goldman, T., \& Perez-Mercader, J. (2016, April 7). Ubiquity of Benford’s law and emergence of the Reciprocal Distribution. Physics Letters A.

https://www.sciencedirect.com/science/article/abs/pii/S0375960116300603 

$\newline$
Sims, E. (n.d.). University of Notre Dame. Eric Sims. 

https://sites.nd.edu/esims/ 
\end{document}